\documentclass[12pt,preprint]{aastex} % CHANGE BEFORE SUBMISSION!!!!
\def\ltsima{$\; \buildrel < \over \sim \;$}
\def\simlt{\lower.5ex\hbox{\ltsima}} % < over ~
\def\gtsima{$\; \buildrel > \over \sim \;$}
\def\simgt{\lower.5ex\hbox{\gtsima}} % > over ~

\begin{document}

\title{Deceleration from Entrainment in the jet of the quasar 1136-135?}

\normalsize \author{F. Tavecchio, L. Maraschi} \affil{INAF -
Osservatorio Astronomico di Brera, via Brera 28, 20121 Milano, Italy}

\author{R.M. Sambruna}\affil{NASA/GSFC, Greenbelt, MD 20771, USA}

\author{M. Gliozzi} \affil{George Mason University,  Fairfax, VA 22030, USA}

\author{C.~C. Cheung\altaffilmark{1}}

\altaffiltext{1}{Jansky Postdoctoral Fellow; National Radio Astronomy
Observatory. Now hosted by Kavli Institute for Particle Astrophysics
and Cosmology, Stanford University, Stanford, CA 94305} 

\affil{MIT, Kavli Institute for Astrophysics \& Space Research, 77
Massachusetts Ave., Cambridge, MA 02139, USA}

\author{J. F. C. Wardle} 
\affil{MS 057, Department of Physics, Brandeis University, Waltham, MA
02454}

\author{C. Megan Urry} \affil{Yale University, Dept. of Astronomy, New
Haven, CT 06520, USA}

\begin{abstract}
Modeling the multiwavelength emission of successive regions in the jet
of the quasar PKS 1136--135 we find indication that the jet suffers
deceleration near its end on a (deprojected) scale of $\sim$400 kpc.
Adopting a continuous flow approximation we discuss the possibility
that the inferred deceleration from a Lorentz factor $\Gamma =6.5$ to
$\Gamma=2.5$ is induced by entrainment of external gas. Some
consequences of this scenario are discussed.
\end{abstract} 


\section{Introduction}

Despite decades of intense efforts, the present knowledge of the
physical processes acting behind the phenomenology shown by
relativistic jets is still rather poor. Basic questions about matter
content, transported power, the role of the magnetic field, the
dissipation mechanisms are awaiting answers (e.g. Blandford
2001). Among these problems, one of the most fundamental concerns the
speed of the flow. While it is widely accepted that the rapid
variability and the extreme compactness displayed by blazars
(e.g. Maraschi \& Tavecchio 2005; Sikora \& Madejski 2001) and the
observation of VLBI superluminal components (e.g. Kellermann et
al. 2004 ) imply relativistic bulk flows (with $\Gamma \sim 10-20$)
for the innermost portions of the jet ($0.1$ pc $ < d < 100$ pc), it is
not clear whether the large scale jets still have relativistic speeds
and how/where deceleration takes place. The present evidence suggests
that FRI jets become trans-relativistic quite early, within few
kiloparsecs (e.g. Bridle \& Perley 1984), while the situation for FRII
jets appears more ambiguous. Mildly relativistic speeds ($\beta <
0.95$) are suggested by studies of high luminosity, lobe dominated
sources based on the jet to lobe prominence (Wardle \& Aaron 1997),
while the interpretation of multiwavelength observations of extended
jets in QSOs points toward highly relativistic speeds ($\Gamma \sim
10$) even at very large scales ($\sim $ 100 kpc; e.g., Tavecchio et
al. 2000, Celotti et al. 2001, Sambruna et al. 2002,2004,
Siemiginowska et al. 2002, Marshall et al. 2005).

Deceleration of FRI jets is commonly believed to be the result of
entrainment of external gas by the flow (e.g. Komissarov 1994,
Bicknell 1994, B94 hereafter). In this framework it is conceivable
that FRII jets, characterized by larger powers, are not substantially
perturbed by entrainment: the fundamental division between FRI and
FRII jets could basically depend on the difference in the mass/energy
flux of the two types of jets (Bicknell 1995, Bowman et al. 1996). An
important unknown parameter is the efficiency with which the jet can
entrain the external gas. Efficient entrainment can be assured by the
onset of fully-developed turbulence at transonic speeds (e.g. B94). In
this scheme, supersonic flows are thought to be less efficient at
collecting material and entrainment could affect only the external
regions of the jet, resulting in the formation of a slow layer
surrounding a fast, unperturbed ``spine'' (e.g. de Young
1986). Nevertheless, even low efficiency entrainment could become
important if it is maintained over very long scales.

The possibility of ``observing''  the gradual slowing-down of a jet could
in principle provide precious information on the physical processes at
work. This approach was successfully developed in great detail for a
few FRI jets where the morphology could be well studied thanks to the
large angular scale (Laing \& Bridle 2002, 2004).  For FR II jets the
stronger asymmetry and smaller angular scale has generally prevented
comparable studies.

Of particular interest are therefore the cases of FRII jets hosted by
intermediate-$z$ QSOs in which multiwavelength observations have
recently shown that the radio to X-ray flux ratio increases dramatically
along the jet. This feature, interpreted in the framework of the
synchrotron-IC/CMB emission model, suggests that the jet undergoes
deceleration. The possible role played by deceleration was first
discussed for the jet of 3C273 (Sambruna et al. 2001). Other cases of
jets showing an increasing radio to X-ray flux ratio were presented in
the survey discussed in Sambruna et al. (2002) and Sambruna et
al. (2004). Georganopoulos \& Kazanas (2004; hereafter GK2004)
proposed an analytic framework to describe the effects of deceleration
on the physical parameters of the jet on the basis of an adiabatic
assumption, remarking that, if the jet decelerates, the magnetic field
as well as the particle energy density will increase due to
compression causing a large increase of the radio synchrotron
emission, while the X-ray flux will decrease because of a reduced
beaming of the CMB radiation field. In Sambruna et al. (2005; Paper I
in the following) we report the results of deep {\it Chandra} and
multiwavelenth observations of PKS~1136-135, which allowed us to image in
detail the emission along the jet.  Modelling the observed
multifrequency emission we derived the profiles of the basic physical
quantities along the jet, which support the deceleration scenario
discussed above. In this paper on the basis of the observational
evidence collected for the jet of 1136-135, already presented in Paper
I, we discuss in more depth deceleration in terms of physical models
and in particular entrainment of external gas.

The plan of the paper is the following: in Sect.2 we derive the
profile of the relevant physical quantities, suggesting deceleration
of the flow. In Sect.3 we discuss the physical origin of the
deceleration. In Sect.4 we discuss in detail the possibility that
deceleration is induced by the loading of the jet through
entrainment. Finally in Sect.5 we conclude. Throughout this work we
use the current ``concordance'' cosmological parameters: $\rm H_0=71\;
Km\; s^{-1}\; Mpc^{-1} $, $\Omega _{\Lambda}=0.73$, $\Omega _m = 0.27$
(Bennett et al. 2003).

\section{Determining the deceleration profile}

One of the most striking features of the the jet of 1136-135 after knot B
is the dramatic increase of its radio luminosity and the associated decay
of its X-ray emission (Fig.1). The brightness profiles directly translate
into the run of the physical quantities found in Paper I by using the
IC/CMB radiative model (Tavecchio et al. 2000; Celotti et
al. 2001). Indeed the increasing radio brightness requires an increase of
one order of magnitude both of the magnetic field and the density of the
radiating electrons (bound together through the equipartition
condition). The decrease shown by the X-ray flux, in turn, implies a
decrease of the Doppler factor, which reduces the amplification suffered
by the target photons of the CMB. Therefore, the observed morphology and
the resulting profiles suggest that the special behavior displayed by
1136-135 could be related to effects induced by the deceleration of the
flow, starting in the region corresponding to knot B-C. Remarkably, the
physical parameters inferred for the hot spot (F in the prsent work, HS
in Paper I) appear to continue the same trend followed by the knots,
supporting the emerging view that the hot spots of jets in powerful QSOs
can be thought of as slow knots, rather than portions of the jet almost
at rest (Georganopoulos \& Kazanas 2003; Tavecchio et al. 2005). We
remark that the morphology and the results of the modelling suggest that
the deceleration is important only after the region B-C. Before that
point the character of the flow seems to be different, with emission
clearly concentrated in knots characterized by a similar speed. In this
paper we only consider the region after knot B, where deceleration seems
to take place.

We review the main features of the model used to fit the data. The
emitting region is assumed to be in motion with a bulk Lorentz factor
$\Gamma$ at an angle $\theta$ with respect to the line of sight. Since
fluxes are extracted within elliptical regions (Paper I), we assume a
volume of the emitting region equal to that of the corresponding
(deprojected, see below) ellipsoid. The emitting region is
homogeneously filled by high-energy electrons, with a power-law energy
distribution $N(\gamma )=K\gamma^{-n}$ ($K$ is the electron
normalization) extending from $\gamma_{\rm min}$ to $\gamma_{\rm
max}$. Electrons emit radiation through the synchrotron and IC/CMB
mechanisms. The low-energy limit of the electron distribution
$\gamma_{\rm min}$ is constrained by the optical and the X-ray
fluxes. The calculations reported in Paper I have been performed
assuming that the observed emission concentrations (knots) mark the
presence of discrete moving components of the jet. In the region of the
jet under consideration here (after knot B) the flow seems to become
more similar to a continuous flow, in which it is difficult to isolate
knots (Paper I). It is likely that the flow is complex, showing the
characteristics of a continuous flow together with the presence of
regions where the dissipation is more efficient, visible as brighter
regions in the jet. Based on this evidence, and to be consistent with
the scenario that we intend to explore in this paper, we recalculated
the physical quantities assuming that, after knot B, the emitting
plasma behaves as a continuous flow, moving in a jet whose pattern is
at rest with respect to the observer. Practically this implies that
the exponent $2+\alpha$ should be used instead of $3+\alpha$ for the
Doppler factor to describe the amplification of the radio flux (Lind
\& Blandford 1985) and that the observed length is deprojected using
the observing angle. The choice of this particular geometry slightly
affects the values of the derived quantities, but does not change the
evidence of a trend suggestive of deceleration.

With the assumptions above it is possible to derive the required
values of the physical quantities by using the observed radio (5 GHZ)
and X-ray (1 keV) fluxes. To break the degeneracy between the model
parameters (we have three physical quantities the magnetic field, $B$,
the electron normalization $K$ and the Doppler factor $\delta$ and two
quantities from observations, the radio and the X-ray flux) it is
necessary to introduce an additional condition. A widely adopted
assumption (although, admittedly, not fully motivated) is that
particles and magnetic field are in equipartition (e.g. Tavecchio et
al. 2004; but see Kataoka and Stawarz 2005). Several evidences
indicate the presence of a proton component in jets (e.g. Sikora \&
Madejski 2000). However, we assume that protons are not directly
coupled to the non-thermal components (relativistic electrons and
magnetic field), although they can contribute to the total
pressure. We then assume equipartition between relativistic electrons
and magnetic field energy densities, $U_B=U_e$, assuming no
contribution from protons. Note that, in any case, the inclusion of
protons into the equipartition condition only slightly affects the
derived quantities (unless protons strongly dominate the total
pressure). It is then possible to find analytical relations providing
a unique set of physical parameters characterizing the emission region
(e.g. Tavecchio et al. 2000). The appropriate analytic relations are
given in Appendix A.

We devoted special attention to the estimate of the uncertainties
affecting the final parameters ($B$, $K$ and $\delta$) deriving from
the uncertainties associated to the input parameters of the model. To
this aim we calculated the ranges covered by $B$, $K$ and $\delta $
 for different values of the input parameters, varying $\gamma
_{\rm min}$ in the range 5--25, the volume of the emitting region in
the range $V-V/2$, the radio and the X-ray fluxes and the radio
spectral slope $\alpha _r$ within the corresponding errors (given in
Paper I). We find that the total uncertainty is largely dominated by
the error on the radio spectral index. We then associate the
total range spanned by $B$, $K$ and $\delta$ for the different input
parameters with the total uncertainty.

A more complex discussion involves the viewing angle, $\theta $.  It
enters as an input parameter in the calculations, through the deprojection
of the observed length. At the same time, the angle is constrained by the
results of the calculations. Indeed, for a derived Doppler factor $\delta
$ (resulting from the modelling), $\theta $ cannot be larger than
$1/\delta$ (see e.g. the discussion in Tavecchio et al. 2004). To fix the
angle we then follow an iterative procedure: we fix an angle, we derive
the parameters, we check if the maximum Doppler factor required by the
knots (including the upper bound deriving from the uncertainties
discussed above) satisfies $\theta < 1/\delta _{\rm max}$. If the
condition is not satisfied we reduce the input viewing angle and we
repeat the procedure, until we reach the maximum angle allowed by the
data, $\theta _{\rm max}=3.8$ deg. All the angles $\theta < \theta _{\rm
max}$ are in principle allowed. However, since $\theta _{\rm max}$ is the
most probable angle and the impact of a different choice is quite small
and $\theta _{\rm max}$ is the most probable angle, we use $\theta _{\rm
max}$ in the modelling.

The profiles of $B$, $K$ and $\Gamma $, derived from the emission
model with equipartition, are shown in Fig. 2.  The value of the
Lorentz factor $\Gamma $ can be derived from $\delta$ once the
observing angle is fixed. We recall that each knot is modelled
separately from the others: therefore the trends we derive are not
by-products of the procedure used, but characteristics of the flow,
and are basically related to the observed trends of radio and X-ray
fluxes along the jet. The evolution of the model parameters with
distance are in qualitative agreement with the adiabatic model
discussed in GK2004. Unfortunately, the large uncertainties prevent a
deeper discussion of this point.

In conclusion our results support the view that the jet of 1136-135
suffers strong deceleration after knot B. The Lorentz factor, $\Gamma$
changes, from (with this specific choice of the angle) 6.5 to 2.4 in a
projected angular distance of $\sim 4$ arcsec, corresponding to a
deprojected linear distance of $\sim 400$ kpc.

\section{Origin of deceleration}

In the following we consider possible deceleration mechanisms, a
question not explicitly addressed in GK2004.  First we estimate the
possible role of the {\it Compton-Drag} effect and then proceed to
discuss entrainment.

\subsection{Compton Drag}

A possible way of decelerating a relativistic plasma is through the
recoil induced by Compton scattering between the electrons present in
the jet and some external radiation field (a process known as {\it
Compton Drag} effect, e.g. Sikora et al. 1996). This mechanism has
been recently proposed as a viable way to decelerate BL Lac jets
within the first parsec from the core (Ghisellini, Tavecchio \&
Chiaberge 2005) and as a source of IR radiation from interknot regions
in large scale jets (Georganopoulos et al. 2005).

In the particular case analysed here the external radiation field is
dominated by the CMB (with energy density $U_{\rm CMB}$), while
contribution of radio emission from the slow part of the jet (beamed
in fast jet frame) is at most comparable to that of the CMB
(e.g. Celotti et al. 2001). A key role in determining the efficiency
of such a process is played by the composition of the jet: clearly,
only when electrons carry a large fraction of the total jet power the
process can affect the jet dynamics. Basically, the equation
describing the evolution of the bulk Lorentz factor with time can be
written as (e.g. Ghisellini et al. 2005):

\begin{equation}
\dot \Gamma \sim \frac{\sigma _T c <\gamma ^2> \Gamma ^2 U_{\rm CMB}}
{m_ec^2\left( \omega \frac{m_p}{m_e} + <\gamma >\right)}
\end{equation}

\noindent
where $\omega = n_p/n_e$. $<\gamma ^2>$ and $<\gamma >$ depend on the
detail of the electron distribution. Using typical values derived
with the IC/CMB model, $n=2.6$, $\gamma _{\rm min}=10$, $\gamma _{\rm
max}=10^5$ (e.g. Sambruna et al. 2004), they can be estimated $<\gamma
^2>\sim 10^4$ and $<\gamma >\sim 25$. We assume the most optimistic case
in which electrons do not substantially cool (or are continuously heated)
and therefore $<\gamma ^2>$ is constant.  The effects of the deceleration
can be estimated through the quantity $r_{\rm dec}=c \Gamma /\dot
\Gamma$, yielding the typical distance at which the deceleration takes
place. Inserting the numerical values we find $r_{\rm dec}\sim 100$ Mpc
%5\times 10^{26} 
for proton-dominated jets ($\omega =1$), $r_{\rm dec}\sim 300 $ kpc for
pair-dominated jets ($\omega =0$) with $<\gamma >\sim 10$. We conclude
that {\it Compton Drag} can be efficient only in the case of a
pair-dominated jet. General arguments applied to blazars (although
not completely conclusive) seem to point toward a proton component for
jets in quasars dominating the dynamics (e.g. Sikora \& Madejski 2000,
Ghisellini \& Celotti 2002a, Maraschi \& Tavecchio 2003; see also
Uchiyama et al. 2005). We can conclude that an important role of the
radiative recoil in determining the inferred deceleration seems unlikely,
although not completely ruled-out.

\subsection{Entrainment}

In the following we discuss the possibility that the inferred decline of
$\Gamma$ along the jet is caused by the continuous matter loading of the
jet through entrainment of external gas entering from the lateral walls
of the jet. This mechanism is believed to be responsible for the early
deceleration of FR I jets (e.g. B94).

Basically, deceleration through entrainment can be understood to
happen through a continuous series of inelastic collisions between the
moving plasma and the external gas at rest (e.g., Icke 1991). As a
result of the collision, part of the kinetic energy is dissipated and
converted into internal energy of the jet, thus increasing the
internal pressure.  We first derive an approximate formula for the
evolution of the Lorentz factor in the simplified case of an
entraining confined mass (cylinder). Although drastically simplified,
this approach allows to identify the key elements of the entrainment
process.

During the motion, the external layer of the cylinder interacts with
the external medium, resulting in a continuous increase of the moving
mass.  Hence a continuous deceleration due to momentum conservation
and an increase of internal energy due the dissipation of kinetic
energy follow. The dynamics of the entraining mass can be described
using the conservation laws for energy and momentum for {\it inelastic}
collisions. The initial mass is $M_o$ with initial Lorentz factor is
$\Gamma _o$. Assuming that the entrained gas is initially at rest,
conservation of energy and momentum yields:

\begin{equation}
\Gamma _0 M_0 + m = \Gamma M = \Gamma \left(
m+\frac{E}{c^2}+M_0 \right)
\label{energy}
\end{equation}
\noindent
\begin{equation}
\beta _o \Gamma _0 M_0 = \beta \Gamma M = \beta \Gamma \left(
m+\frac{E}{c^2}+M_0 \right)
\label{momentum}
\end{equation}

\noindent
where we have indicated with $m$ the mass of the material entrained by
the blob and $E$ is the internal energy accumulated in the collision.
Eqs.(\ref{energy})-(\ref{momentum}) give the Lorentz factor $\Gamma $
as a function of the rest-mass $m$ of the collected material:

\begin{equation}
\Gamma = \frac{a+\Gamma _o}{(a^2+2a\Gamma _o +1)^{1/2}}
\label{m}
\end{equation}

\noindent
where $a=m/M_o$. The Lorentz factor reached depends only on the ratio
between the total entrained and initial mass of the blob. For
$m/M_o<<1/\Gamma _o$ the Lorentz factor remains almost constant, while
for $m/M_o>>1/\Gamma _o$ $\Gamma$ tends to 1. Thus the critical scale for
deceleration is reached when the collected mass is of order $m_{\rm
crit}\sim M_o/\Gamma _o$. To halve the initial bulk Lorentz factor
$\Gamma_o$ the collected mass must be $m= 1.5 M_0/\Gamma _o$. In the
limit of the inelastic assumption the collected mass needed to slow down
a highly relativistic mass is smaller for higher $\Gamma_o$.

In the following we apply the rigorous hydrodynamical treatment
developed by B94 to describe the deceleration of the jet of 1136-135
in order to discuss the plausibility of the entrainment mechanism for
this particular case.  Laing \& Bridle (2002) used the same approach
to model the dynamics of the jet of the FRI galaxy 3C31. Our analysis
will be not as detailed, due to the more limited information allowed
by the larger distance and smaller angular scale. The treatment
assumes that the system is adiabatic, the dissipated kinetic energy
being stored in the plasma internal pressure. This condition is
satisfied in the case under study, for which it is known that the
radiative efficiency is quite low (e.g. Tavecchio et al. 2000;
Sambruna et al. 2002).

We use momentum and energy conservation in the form given by B94 to
follow the evolution of the decelerating flow. An important parameter is
$\chi = \rho c^2/(\epsilon +p)$, representing the ratio between the rest
mass density of the particles within the jet and the sum of internal
energy density and pressure (enthalpy). Basically this parameter measures
the ratio between the rest-mass energy and the (random) kinetic energy of
the particles of the jet, as measured in the comoving frame of the
flowing plasma. Values of $\chi >1$ indicate cases for which the average
kinetic energy of the particles is less than the rest mass-energy, i.e. a
non relativistic plasma. On the contrary values of $\chi <1$ characterize
a relativistic plasma.  Note that, based on the modelling of the
observed emission, we can directly probe only the non-thermal component
of the plasma (relativistic electrons and magnetic field). However, the
plasma can also contain non relativistic or mildly relativistic electrons
and protons (the ``thermal'' component), contributing to (and possibly
dominating) the mass density and the total pressure.  Defining the
initial internal pressure of the jet $p_0$, the initial Lorentz factor
$\Gamma _0$, the initial cross sectional area of the jet $A_0$, the
initial $\chi _0$ parameter and the corresponding downstream parameters,
the coupled expressions for momentum and energy conservation read:

\begin{equation}
(1+\chi)\Gamma^2\beta^2+\frac{1}{4}= \left[(1+\chi _0)\Gamma^2_0
\beta^2_0+\frac{1}{4}\right]\left(\frac{p_0 A_0}{p A} \right)
\label{mom}
\end{equation}
\noindent
and:
\begin{equation}
\left[(\Gamma -1)\chi +\Gamma \right]\Gamma \beta=
\left[(\Gamma_0-1)\chi _0+\Gamma_0 \right]\Gamma_0 \beta_0 
\left(\frac{p_0A_0}{p A} \right)
\label{en}
\end{equation}

B94 assumed that the dominant role in determining the pressure is played
by relativistic particles for which the relativistic equation of state
$p=\epsilon /3$ can be applied; we adopt the same assumption. As can be
verified {\it a posteriori}, the assumption is fully consistent with the
results only at the end of the deceleration process. However, we verified
that even in the unrealistic case of a non-relativistic equation of state
there are just small corrections to the results (see also Laing \& Bridle
2002) and the general conclusions of our study do not change. 
Analogously, for simplicity, in Eq.(\ref{mom}) the terms corresponding to
the effects of the external medium (expected to be small) have been
dropped.

To solve the equation above we assume that the cross sectional area of
the jet, $A$, remains approximatively constant from C to F. Although this
assumption is not fully consistent with our choice to neglect the effects
of the external medium, the ``cilindrical'' approximation seems tolerable
since the expansion factor in the considered region is at most a factor
of 2. Moreover the jet could be self-confined by, for instance, a large
scale magnetic field.

Once the initial state is defined, the coupled non-linear equations can
be solved to provide $p$ and $\chi $ at each value of $\Gamma $,
considered as a parameter. Other important derived quantities, such as
the comoving density $\rho $ and the entrainment rate, can also be
calculated.

\section{Results}

The basic features of the deceleration process are shown in Fig. 3
through the evolution of the parameter $\chi $, the density and the
pressure as a function of the Lorentz factor reached by the flow. For
this calculation we assume initial conditions corresponding to the
parameters found above for knot C of 1136-135, after which the
deceleration is more evident (Fig. 2). As discussed above, we do
not have any information about the thermal components of the
plasma. The mass density has been calculated assuming a composition of
1 cold proton per emitting electron. For the pressure at knot C, we
assume two cases. In the first, we assume that the pressure is
dominated by the non-thermal components, $p_o=p_e+p_B$ (solid
lines). In the second case (dashed lines) we assume a pressure ten
times that the non-thermal one, thought to be provided by a thermal
component.

To satisfy energy and momentum conservation, each small entrained mass
induces a small deceleration of the flow and a small increase of the
internal energy (assumed to be shared among all the protons inside the
jet). This produces variations of the $\chi $ parameter, as shown in
Fig. 3 (top). Assuming a jet initially dominated by cold gas $\chi >>1$,
the dissipation induced by the continuous loading implies a decrease of
$\chi $, because the pressure increases faster than the density of the
jet. When the jet is almost completely stopped $\chi $ is of order unity
and starts to increase. Note that the regime considered by Laing \&
Bridle (2002) for 3C31 corresponds to low $\Gamma$, corresponding to the
region where $\chi $ increases during the deceleration. In our
application to a highly relativistic FRII jet, we are considering the
regime where $\chi $ is large at the beginning and decreases along the
jet.

The profile of the comoving density (normalized to the initial value)
is shown in Fig.3 (middle).  Also shown in the same figure is the
profile one would get in the case of deceleration without entrainment
(with the same geometry used for the other calculations): in
this case the increase of the density only reflects the effect of
deceleration and consequent compression of the flow (following the law
$\rho/\rho_0=\Gamma _0/\Gamma $ expressing the conservation of the
number of particles). The difference between this curve and the other
curves represents the contribution to the total density of the
entrained material accumulated along the path. Comparing the densities
with and without entrainment it can be seen from Fig. 3 that the
entrained mass (proportional to the density difference) needed to
decelerate the jet from $\Gamma _0\sim 6.5$ to $\Gamma \sim 3.2 $
(halving the initial Lorentz factor) represents a small fraction
(about 10\%) of the original rest-mass.

In the bottom panel of Fig.3 we report the pressure of the plasma
for different Lorentz factors reached by the flow. We also plot
(points) the values of the pressure supported by the non-thermal
components only (magnetic field and non-thermal electrons), as
estimated from the modelling of the emission. The errorbars along the
$x-$axis indicate the uncertainty associated to the Lorentz factor of
each region, as reported in Fig.2. Clearly the non-thermal component
alone can account for a small fraction (around 10\%) of the pressure
required by the conservation laws, even in the case in which the
initial pressure is completely non-thermal. Therefore much of the
pressure is provided by the energy stored within the invisible thermal
component.

The ability of the entrainment process to decelerate the jet is
related to the quantity of matter effectively entrained by the jet
along its path, expressed by the entrainment rate. The essential
parameter characterizing this quantity is the entrainment
velocity. Attributing the ``observed'' deceleration to entrainment we
can derive the required entrainment rate and entrainment velocity in
the specific case of 1136-135.

The entrainment rate can be defined starting from the law
describing the evolution of the jet mass flux (see Eq. 9 of B94):
\begin{equation}
%\Gamma _2 \rho _2 \beta _2c A_2 = \Gamma _1\rho _1\beta _1c A _1+
%\dot{M}_{\rm ent} \,\, \rightarrow \,\,
\Gamma _2 \rho _2 \beta _2c A_2 = \Gamma _1\rho _1\beta _1c A _1+  
%\dot{M}_{\rm ent}
\int _1^2 \rho _{\rm ext} v_{\rm ent} dA
\label{vent}
\end{equation}
\noindent
where subscripts 1 and 2 indicate quantities of the jet measured at
distances $x_1$ and $x_2$, $dA\simeq 2\pi r_j dx$ is the jet lateral
surface and $v_{\rm ent}$ is an effective entrainment speed.  The
(average) entrainment rate is thus defined as:
\begin{equation}
\frac{\Gamma _2 \rho _2 \beta _2c A_2 - \Gamma _1\rho _1\beta _1c A _1}
{\Delta x_{1,2}} \, = \, \frac{1}{\Delta x_{1,2}} \int _1^2 \rho _{\rm ext}
 v_{\rm ent} dA
\end{equation}

The average entrainment rate calculated from C to F (comparable
values can obtained for the intermediate regions C-D, D-E, E-F) is
$\sim 2\times 10^{24}$ g s$^{-1}$ kpc$^{-1}$ (the distance has been
deprojected assuming $\theta=3.8$ deg; the inferred entrainment
rate would scale as $\propto \theta $). For comparison, this value is
about an order of magnitude larger than the level required for the
deceleration of the jet of 3C31 (Laing \& Bridle 2002). In the latter
case the collecting surface is smaller and the density much
higher. Further assuming that the external density and the entrainment
velocity do not change dramatically from C to F and specifying the
external density, it is possible to derive the entrainment speed
$v_{\rm ent}$. Assuming that in the region from C to F the external
gas has a density $n_{\rm ext}=10^{-5}$ cm$^{-3}$ (see below) and that
the jet in the region considered here has a radius of $r=10^{23}$ cm
we find $v_{\rm ent} = 4 \times 10^7$ cm/s. The corresponding value
for the jet of 3C31 lies around $v_{\rm ent} = 10^5-10^6$
cm/s. Therefore, the entrainment process in 1136-135 appears to be
more efficient than that occurring in the low-power jet of 3C31.

\subsection{The onset of deceleration}

A question naturally arising is whether the parameters required by the
entrainment model for the "observed" deceleration in the region C--F
are consistent with the propagation of the jet from the nucleus to the
large scale. For this scope, we consider the simplest hypothesis that
entrainment is active with the same efficiency (measured by the
parameter $v_{\rm ent}$) all along the path of the jet and assume a
conical, entraining jet\footnote{In the previous section the jet
radius has been approximated as constant in the region from
C--F. However, in the following we want to study the cumulative effect
of entrainment from very small to large scales, and thus the increase
of the jet radius with the distance must be taken into account.}.

It is possible to check that the critical ``cumulative'' entrained mass
flux for substantial deceleration in the case of a continuous jet
satisfies a relation analogous to that derived above for the case of a
blob, $m_{\rm crit}=M_o/\Gamma _o$.  Deceleration becomes important
when the contribution of the cumulative mass entrained by a jet slice
along its path is of order $1/\Gamma $ of the jet slice mass.  This
condition can be put on an explicit form:

\begin{equation}
\int _1^2 \rho _{\rm ext} v_{\rm ent} dA \sim \frac{F_{M,1}}{\Gamma
_1} \, = \, \rho _1\beta _1c A_1
\label{fent}
\end{equation}

In order to estimate the entrained mass flux (Eq.\ref{fent}) we need
to specify the distribution of the external gas density, $\rho _{\rm
ext} \simeq n _{\rm ext} m_p$ while $v_{ent}$ will be assumed to be
constant. From the {\it Chandra} image we infer the presence of faint
diffuse emission surrounding the QSO. The distribution of the gas can
be modelled as a King profile with core radius $x_o=40$ kpc and
central density $n_o= 5\times 10^{-4}$ cm$^{-3}$ (Sambruna et al., in
prep).  This result is consistent with other X-ray observations which
show that, in general, nearby ($z<1$) QSOs lie in environments typical
of small groups (Hardcastle \& Worrall 1999; Crawford \& Fabian 2003).
At distances of interest here the density can be well modelled as a
power-law:

\begin{equation}
n(x)=n_o\left( \frac{x}{x_o}\right)^{-1.5} 
\end{equation}

\noindent
with $x_o=40 $ kpc, $n_o=3\times 10^{-4}$ cm$^{-3}$. 
From the profile above, the density in the region from knot C to F is
expected to be of the order of $n_{\rm ext}\sim 10^{-5}$ cm$^{-3}$, which
has been used above to derive the value of $v_{\rm ent}\sim 10^8$
cm/s. Assuming that $v_{\rm ent}$ does not substantially change along
the jet, it is thus possible to estimate the total matter collected by
the jet from the kpc scale to the hundreds of kpc scale, where
deceleration becomes effective. Assuming a conical jet, the area $dA$
can be expressed as $dA=2\pi r dx$, with $r=r_o x/x_o$, where $r_o$ is
the radius at the initial point $x_o$ (typically it is assumed
$r_o/x_o\sim 0.1$). 

Even if the entrainment efficiency is constant and the density of the
external medium is decreasing, the entrained mass flux steadily
increases, due to the continuous increase of the jet's
surface. Calculating the integral in Eq. (\ref{fent}) we find, as
expected, that the typical scale at which deceleration is effective is
$x_1\sim 100$ kpc. Thus the propagation of the jet up to large
distances can be understood. The terminal part of the jet of 1136-135
can be identified as the region where the entrained mass becomes
critical causing substantial deceleration. We note however that in the
schematic model we have drawn with constant entrainment speed, the
deceleration profile is very regular. Clearly there must be additional
conditions (e.g. the onset of strong instabilities) which enhance
the process at particular sites driving the localized occurrence of
deceleration.

\section{Discussion and Conclusions}

The proposed scenario can explain in a plausible way the deceleration
inferred for the jet of 1136-135 (and possibly in the other sources
showing the same radio-to-X-rays increasing trend, GK2004).  A more
detailed understanding of the processes at work and the comparison
with the observed properties of jets necessarily involves several
still poorly-known physical issues.

It is worth remarking that the evidence for deceleration in the jet of
1136-135 is based on the observed run of radio and X-ray fluxes along the
jet and on the use of a specific emission model to interpret the
multiwavelength observations (for a discussion of alternative emission
models see Paper I). In this respect, the conclusion about deceleration
is not completely model-independent. Other specific assumptions made in
the modelling, such as the equipartion condition, affect (although in a
minor way) the conclusion. Moreover, apart from the application of a
specific emission model, one has to bear in mind that the result is based
on the hypothesis that the emission comes from a not-structured flow. At
least part of the increase of the radio flux close to the terminal part
of the jet could be due to the contribution from another component
(e.g. the plasma backflowing after the compression at the hot-spot).

The approach that we used to study the deceleration is based on a pure
hydrodynamical treatment.  The contribution of all the components of
the flow (protons, electrons, magnetic field) is included in just one
parameter, the total pressure. Likely, much of the energy dissipated
during the deceleration will be stored in the heated proton
component. To follow the evolution of the other components (thermal
and non-thermal electrons, magnetic field) we should specify the
actual mechanisms coupling each component to the others. The
determination of the coupling between protons, magnetic field and
electrons, determining the fraction of dissipated energy that goes
into the magnetic and non-thermal electron components, is a
longstanding problem in astrophysics, clearly beyond the scope of this
work. The derived profiles are in agreement with the possibility that
the increase of the magnetic field and of the density of the
relativistic electrons just follows the adiabatic heating induced by
the deceleration. However, given the large uncertainties, this
evidence is not conclusive. The contribution of non-thermal electrons
and magnetic field to the total pressure is small, around
10\%. Therefore, a large portion of the dissipated kinetic energy
resides in the hot protons and is not radiated away.

Another interesting point is related to the nature of the mechanism
able to induce the assumed entrainment. It is widely assumed that
turbulence can effectively work only in slow (transonic or subsonic)
jets.  Apart from fully developed turbulence, it is also possible that
entrainment is mediated by viscosity effects induced by small-scale
turbulence.  It is worth noting that in the calculation we have
presented it is supposed that the entire jet is decelerated under the
effect of the entrainment. More realistically, these processes will
induce a radial structure in the jet, with the development of a
sheared flow (e.g. De Young 1996).

In the simple scheme presented here, the jet starts to be decelerated
when the amount of collected gas is of the order of 1/$\Gamma $ of the
mass transported by the jet. In this framework, jets characterized by
different mass fluxes will experience different behaviours. Large mass
fluxes will assure that, under the same conditions of external gas
density and entrainment rate, the jet will reach its hotspot almost
unperturbed. On the other hand, jets with a small mass flux will be
decelerated soon. It is tempting to further speculate along these lines,
connecting the deceleration of the jet to other properties of the central
AGN. To this aim we indicate the jet mass flux as $\dot{M}_{\rm out}$,
and we note that $\dot{M}_{\rm out}$ is the mass output rate of the
central engine, which is probably some fraction of the accretion mass
rate $\dot{M}_{\rm out}\propto \dot{M}_{\rm acc}$ (e.g. Ghisellini \&
Celotti 2002b). From the condition expressed by Eq.(\ref{fent}) and the
definitions above we can find that the deceleration length $x_1$ is
%\begin{equation}
$x_1 \propto \dot{M}_{\rm acc}/r_o$,
%\end{equation}
%\noindent
(the exact dependence is related to the jet and the external density
profile) where we have supposed that all the jet, before the
deceleration, have a similar bulk Lorentz factor. If the initial
radius of the jet scales as the gravitational radius of the central
black hole $r_o \propto M_{BH}$, we finally find $x_1 \propto
\dot{m} _{\rm acc}$, where $\dot{m} _{\rm acc}$ is the accretion rate
in Eddington units. Thus it is possible to identify sources for which
the jet is effectively decelerated within short scales with sources
with low $\dot{m} _{\rm acc}$ and {\it viceversa}, a situation
reminiscent of the FRI/FRII division.

\acknowledgements

We thank Gabriele Ghisellini for useful discussions and the anonymous
referee for constructive comments and criticisms. F.T and L.M
acknowledge support from grant COFIN-2004-023189-005.

\appendix
\section{Analytical estimate of the physical parameters}

\noindent
$\bullet $ {\bf Synchrotron EC/CMB emission.} 
Assuming a simple power-law distribution for the emitting electrons:
\begin{equation}
N(\gamma )=K\gamma ^{-n} ,
\end{equation}
\noindent
the synchrotron luminosity can be written as (e.g. Ghisellini et
al. 1985):

\begin{equation}
L_s(\nu _s)=c(\alpha ) K B^{1+\alpha } V \delta ^p\nu _s^{-\alpha }
\label{lsincch}
\end{equation}
\noindent
where $V$ is the volume of the emitting region, $\alpha =(n-1/2)$ is
the spectral index, $c(\alpha )$ is a constant and the luminosity is
calculated at the frequency (observer frame) $\nu _s$. The exponent
$p$ of the Doppler factor depends on the geometry of the flow:
$p=3+\alpha $ for a moving ``blob'', $p=2+\alpha $ for a flow whose
pattern is fixed in the observer frame (Lind \& Blandford 1985).

To express the luminosity emitted through the IC/CMB ee start from the
the standard relation for the inverse Compton luminosity:

\begin{equation}
L_c(\nu )=\frac{4}{3}\sigma _Tc V U_{\rm rad} \delta ^{p_1} \int _{\gamma _{\rm
min}}^{\gamma _{\rm max}} \gamma ^2 N(\gamma ) \delta
(\nu - \nu _c) d\gamma
\label{lc}
\end{equation}
\noindent
where $p_1=4+2\alpha $ for a ``blob'' (Dermer 1995) and $p_1=3+2\alpha
$ for a continuous flow. The above expression assumes that the
spectrum of the soft radiation is extremely peaked around the
frequency $\nu _o$ and then $\gamma $ and $\nu _c$ can be related by:
\begin{equation}
\gamma=\left( \frac{\nu _c}{\nu _o} \right)^{1/2}.
\end{equation}
\noindent
Using this expression we can rewrite eq. (\ref{lc}) as:
\begin{equation}
L_c(\nu _c)=\frac{2}{3}\sigma _Tc V U_{\rm rad} K \nu ^{\alpha -1}_{o} \nu
_c^{-\alpha} \delta ^{p_1} 
\end{equation}
\noindent
Using (\ref{lsincch}) to eliminate $K$, we obtain:

\begin{equation}
\delta = B \eta 
\label{deltacmb}
\end{equation}

\noindent
where the factor $\eta$ is:

\begin{equation}
\eta = \left[ \frac{3}{2} \frac{c(\alpha)}{\sigma _Tc} \frac{1}{U_{\rm
rad} \nu ^{\alpha -1}_{o}} \frac{L_c(\nu _c)}{L_s(\nu _s)}\left(
\frac{\nu _s}{\nu _c} \right)^{-\alpha } \right]^{1/(1+\alpha )}.
\end{equation}\\

\noindent
$\bullet $ {\bf Equipartition condition}.

If we assume equipartition ($U_B=U_e$) between the magnetic energy
density $U_B$ and the energy density of relativistic electrons:

\begin{equation}
U_e=mc^2\int _{\gamma _{\rm min}} ^{\gamma _{\rm max}}N(\gamma )\gamma d\gamma.
\label{distch} 
\end{equation}
\noindent
we obtain the following relation between $\delta $ and $B$:

\begin{equation}
\delta =\xi B^{-(3+\alpha)/p}
\label{deltaeq}
\end{equation}

\noindent
where $\xi$ is:

\begin{equation}
\xi=\left[\frac{8\pi m c^2L_s(\nu _s) f}{c(\alpha)\nu _s^{-\alpha
}V}\right]^{1/p}.
\end{equation}

\noindent
The factor $f$ depends on the details of the electron distribution

\begin{equation}
f=\frac{1}{n-2} \left(\gamma _{\rm min} ^{2-n}-\gamma_{\rm max}^{2-n}
\right). 
%- \frac{1}{n-1} \left(\gamma _{\rm min} ^{1-n}-{\rm \gamma_{\rm
%max}^{1-n} \right).
\end{equation}

Equating Eq.(\ref{deltaeq}) and Eq. (\ref{deltacmb}) we can write the
magnetic field as:
\begin{equation}
B=\left( \frac{\xi}{\eta}\right)^{p/(p+3+\alpha)}
\end{equation}
and, inserting in Eq. (\ref{deltacmb})
\begin{equation}
\delta=\eta \left( \frac{\xi}{\eta}\right)^{p/(p+3+\alpha)}
\end{equation}

The normalization $K$ of the emitting electrons can be derived
rearranging the expression of the synchrotron emission,
Eq.(\ref{lsincch}).\\

\newpage

\vskip 1.5 truecm

\centerline{ \bf Figure Captions}

\vskip 1 truecm

\figcaption[1]{Profile of the X-ray (1 keV) and radio (4.9 GHz) flux along the
jet of PKS 1136--135 (data taken from Sambruna et al. 2005).\label{f1}}

\figcaption[2]{Profiles of the relevant quantities ($\Gamma$, top panel, $B$
and $K$, lower panel) for regions B--F as estimated from the radiative
model.\label{f2}}

\figcaption[3]{The parameter $\chi$ (top), the comoving density
(normalized at the initial value, middle) and the pressure (in units
of $10^{-12}$ dyn/cm$^{-2}$) as a function of the Lorentz factor of
the jet, calculated assuming the initial conditions inferred for knot
C of the jet of 1136-135. Solid lines refer to the case of an
initial pressure equal to that of magnetic field and relativistic
electrons, while dashed lines report the results assuming a pressure
ten times that the non-thermal one. The dotted line in the density
panel visualizes the value of the density in the case of deceleration
without entrainment.  Points overimposed on the pressure curve indicate
the value of the pressure provided by magnetic field and non-thermal
electrons derived for the different portions of the flow through the
modelling of the emission (Fig.2).\label{f3}}

%%%%%%%%%%%%% FIGURES %%%%%%%%%%%%%%%%%%%

\begin{figure}[]
\noindent{\plotone{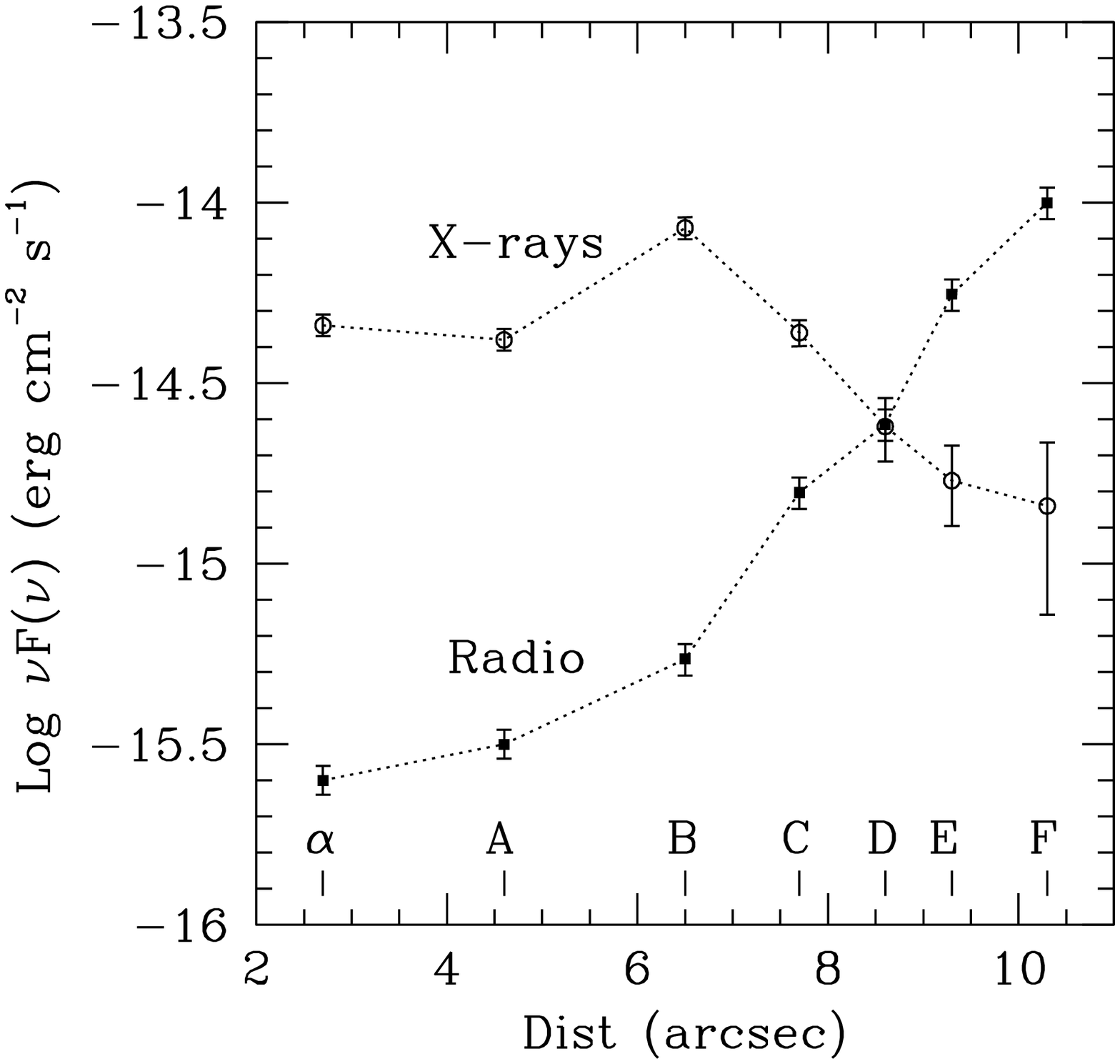}}
%\caption{Profile of the X-ray (1 keV) and radio (4.9 GHz) flux along the
%jet of PKS 1136--135 (data taken from Sambruna et al. 2005).}
%\label{profile}
\end{figure}

\begin{figure}[]
\noindent{\plotone{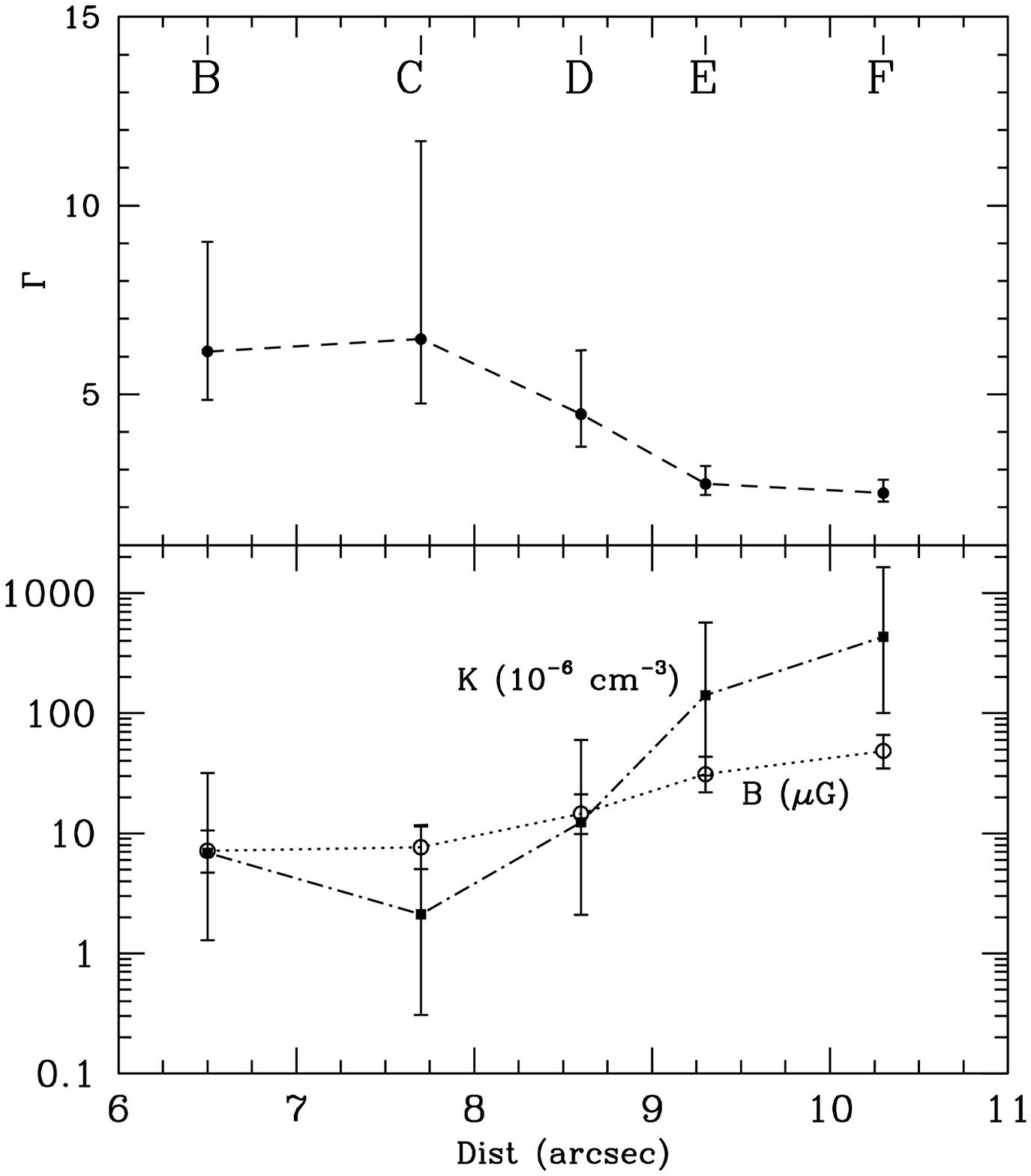}}
%\caption{Profiles of the relevant quantities ($\Gamma$, top panel, $B$
%and $K$, lower panel) for regions B--F as estimated from the radiative
%model.}
%\label{gkcfr}
\end{figure}

\begin{figure}[]
\noindent{\plotone{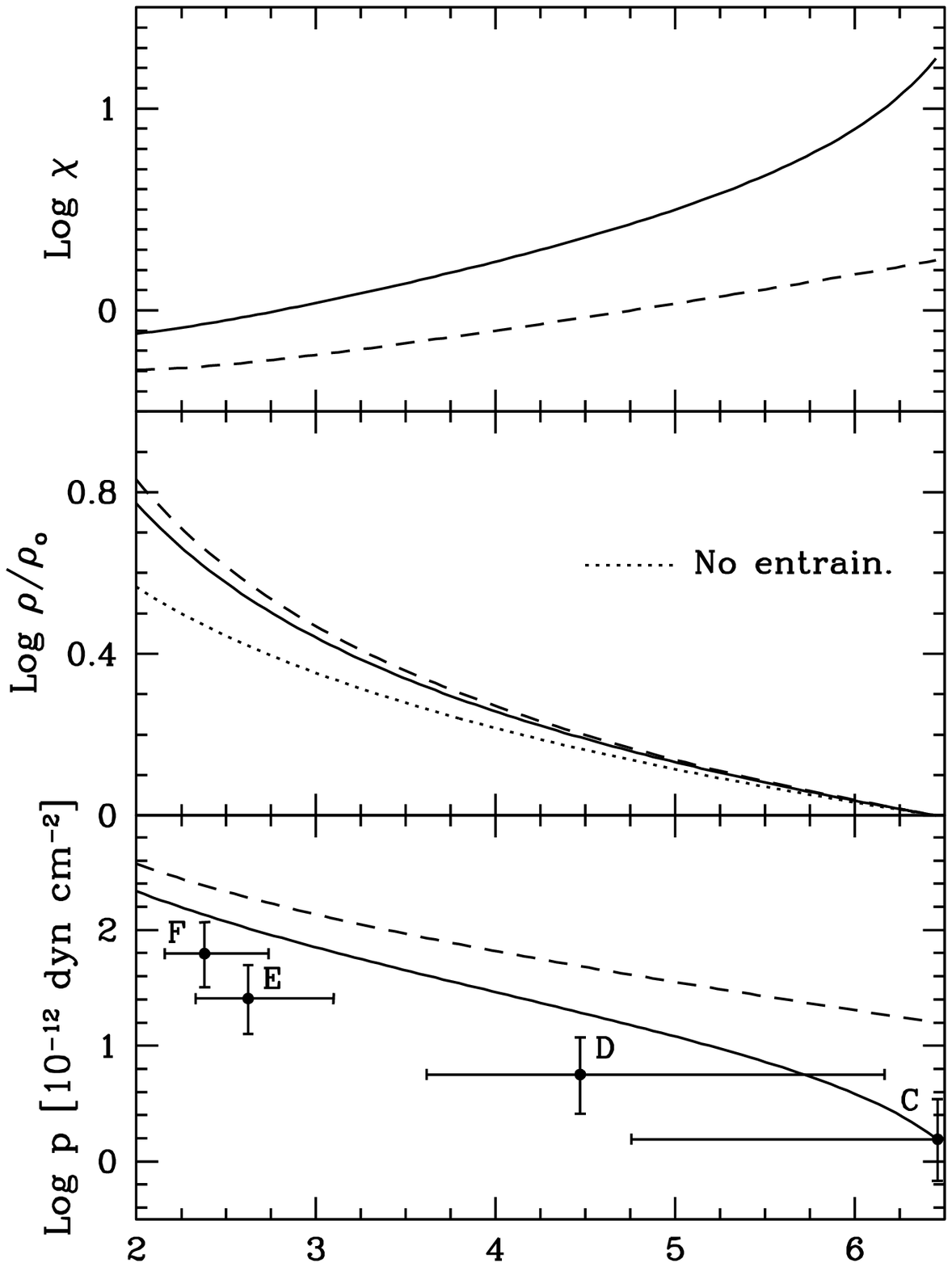}}
%\caption{The parameter $\chi$ (top), the comoving density (normalized at
%the initial value, middle) and the pressure (normalized at the initial
%value, bottom) as a function of the Lorentz factor of the jet,
%calculated assuming the initial conditions inferred for knot C of the
%jet of 1136-135. Solid line and dashed line report the free and
%confined jet case, respectively. The dotted line in the density panel
%visualizes the value of the density in the case of deceleration without
%entrainment.  Points overimposed on the pressure curve indicate the
%value corresponding to the Lorentz factors derived for the different
%portions of the flow through the modelling of the emission.}
%\label{bick}
\end{figure}

\end{document}